\documentclass[prd,amsmath,floats,amssymb, floatfix,superscriptaddress,nofootinbib,onecolumn]{revtex4} 

\usepackage[plainpages=false, colorlinks=true, anchorcolor=blue, linkcolor=blue, citecolor=blue, bookmarks=false]{hyperref}
\usepackage[T1]{fontenc}
\usepackage{graphicx}
\usepackage{dcolumn}
\usepackage{bm}
\usepackage{longtable}
\usepackage{adjustbox} 
\usepackage{multirow}
\usepackage{booktabs}

\usepackage{adjustbox}
\usepackage{float}
\usepackage[caption=false]{subfig}
\usepackage[paperwidth=210mm,paperheight=297mm,centering,hmargin=2cm,vmargin=2.5cm]{geometry}

\begin{document}
\include{notations}
\preprint{APS/123-QED}

\title{Search for Dark Matter Annihilation to gamma-rays  from SPT-SZ selected Galaxy Clusters}

\author{Siddhant Manna}
 \altaffiliation{Email:ph22resch11006@iith.ac.in}
\author{Shantanu Desai}
 \altaffiliation{Email:shntn05@gmail.com}
\affiliation{
 Department of Physics, IIT Hyderabad Kandi, Telangana 502284,  India}





\begin{abstract}
We search for dark matter annihilation from galaxy clusters in the energy range from 1-300 GeV using nearly 16 years of Fermi-LAT data. For this purpose, we use 350 galaxy clusters selected from
the 2500 $\rm{deg^2}$ SPT-SZ  survey. We model the dark matter distribution using the NFW profile for the main halo along with the Einasto profile for the substructure. The largest signal is seen for the cluster SPT-CL J2021-5257 with a significance of around $3\sigma$. The best-fit dark matter mass and annihilation cross-section for this cluster are equal to $(60.0 \pm 11.8)$ GeV and $\langle \sigma v \rangle= (6.0 \pm 0.6) \times 10^{-25} \rm{cm^3 s^{-1}}$  for the $\bar{b} b$ annihilation channel. However, this central estimate is in conflict with the limits on annihilation cross-section from dwarf spheroidal galaxies, and hence cannot be attributed to dark matter annihilation. Three other clusters show significance between $2-2.5\sigma$, whereas all the remaining  clusters show null results. The most stringent 95\% c.l. upper limit for the WIMP annihilation cross-section among all the clusters is from SPT-CL J0455-4159, viz. $\langle \sigma v \rangle  = 6.44 \times 10^{-26} \text{cm}^3 \text{s}^{-1}$ for $m_{\chi} = 10$ GeV and $b\bar{b}$ annihilation channel. 

 \end{abstract}


\maketitle
\section{\label{sec:level1}Introduction\protect}
The dark matter problem is one of the most outstanding unsolved problems in Astrophysics~\cite{Jungman,Bertone2005,BertoneHooper,Bosma}. Although 
a large number of observations have pointed to a concordance model of the universe, which consists of about 25\% cold dark matter~\cite{Planck20}, the identity of the cold dark matter candidate is unknown. One of the most well motivated and  widely studied dark matter candidate is a weakly interacting massive particle (WIMP), which is a hypothetical stable particle, assumed to be a Majorana fermion  with masses in the GeV-TeV range and weak-scale interactions~\cite{Weinberg77,Jungman,Bertone2005}. The most well motivated  WIMP candidate is the lightest supersymmetric particle, which is usually the neutralino~\cite{Ellis84}.
Such a hypothetical particle has velocity averaged  cross-section given by  $ \langle \sigma v \rangle \approx   3 \times 10^{-26} \rm{cm^{3} sec^{-1}}$. This cross-section yields   the correct relic abundance,  and enables the WIMP to  decouple from the rest of the universe at non-relativistic velocities, thereby providing all the right properties needed for a cold dark matter candidate~\cite{Jungman,Bertone2005,Dasgupta}. 
There are three kinds of experiments designed to look for signatures of WIMPs. The first type of search involves direct detection of WIMPs, where one looks for signatures of WIMP scattering with nuclei in an underground experiment~\cite{Schumann}. The second search is referred to as indirect detection~\cite{Zeldovich80,Cirelli,Donato14}, where one looks for signatures of WIMP annihilation into secondary particles such as such as neutrinos~\cite{Desai04}, positrons~\cite{Cholis}, anti-protons~\cite{Heisig}, radio waves~\cite{Chan}, and   gamma-rays~\cite{Funk2015}. Finally, one can also look for dark matter signatures at particle colliders such as LHC~\cite{colliders}.

In this work, we search for dark matter annihilation to gamma-rays using galaxy clusters. Galaxy clusters  are  the most massive  gravitationally bound and virialized structures in the Universe and act as a unique laboratory to probe cosmology~\cite{Kravtsov2012, Allen2011, Vikhlininrev} and fundamental Physics~\cite{Bohringer16,Desai18, Boraalpha, BoraDesaiCDDR, BoraDM}. Most of the mass of galaxy clusters (80-85\%) is made up of dark matter~\cite{Allen2011}. Therefore, galaxy clusters are the largest reservoirs of dark matter and hence provide excellent targets  in looking for signatures of  dark matter annihilation~\cite{Profumo05}.

The  Fermi Gamma-ray Space Telescope which  was launched in 2008 is currently the most sensitive telescope for  gamma-ray astronomy at GeV energies. The  Large Area Telescope (LAT) is one of the two instruments onboard this detector. Fermi-LAT is sensitive to high energy gamma rays from various astrophysical sources. It is a pair-conversion telescope that is sensitive to photons  between the energy range of 20 MeV to more than 300 GeV~\cite{Atwood2009} and is sensitive to dark matter annihilation if the dark matter is a WIMP with masses in the GeV range, since the photon energy from the annihilation products should be within the detector sensitivity.

A large number of searches have been done looking for dark matter annihilation in galaxy clusters using Fermi-LAT~\cite{Ackermann2010,Ando12,Huang11,Han2012,Hektor,Anderson16,Liang16,FermiVIRGO,Lisanti18,Thorpemorgan2021,Shen2021,DiMauro2023,Song2024}. We briefly summarize these works.
The very first search for dark matter annihilation from clusters was done using 11 months of Fermi-LAT data  from six X-ray selected clusters in the HIFLUGCS catalog~\cite{Ackermann2010}. No significant gamma-ray detection was reported in this analysis. Subsequently, null results were reported based on  a search from 49 galaxy clusters located at high galactic latitude using 2.8 years of Fermi-LAT data~\cite{Ando12}. The most stringent upper limit from this analysis was from the Fornax cluster, given by  $\langle \sigma v \rangle \lesssim  (2-3) \times 10^{-25}~\rm{cm^3 s^{-1}}$ for a  WIMP mass  of around 10 GeV~\cite{Ando12}. Then, another search for dark matter annihilation and decay was done from eight nearby galaxy clusters (both individually and from a stacked analysis) using three years of Fermi-LAT data~\cite{Huang11}.  No evidence for gamma-ray emission  at $3\sigma$ significance due to dark matter decay or annihilation  was found from this analysis. Limits on the annihilation cross-section of around  $10^{-25}~\rm{cm^3 s^{-1}}$ were then obtained~\cite{Huang11}.  A similar search for dark matter annihilation from three nearby clusters, namely,  VIRGO, Coma, Fornax  was done using 45 months of Fermi-LAT data~\cite{Han2012}. After removing the galactic and extragalactic gamma-ray foregrounds, a residual diffuse emission was found, which was attributed to additional point sources not included in the Fermi-LAT two-year catalog~\cite{Han2012}. 
After accounting for these, no significant extended emission was detected  from the aforementioned  three clusters,  and upper limits on the cross section for dark matter annihilation were obtained, which were more stringent than those obtained from Milky way dwarf galaxies~\cite{Han2012}. Subsequently, ~\citet{Hektor}  found evidence for a double-peaked line at 110 and 130 GeV from 18 nearby brightest galaxy clusters with more than $3\sigma$ significance using 218 weeks of Fermi-LAT data, which they attributed to dark matter annihilation. However, this signal could not be confirmed in a subsequent follow-up analysis using enhanced exposure~\cite{Anderson16,Liang16}.
A dedicated search for dark matter  annihilation and point source emission from the VIRGO cluster was undertaken by the Fermi-LAT collaboration using three years of data, which reported null results~\cite{FermiVIRGO}. This work then calculated limits on velocity averaged DM annihilation cross-section. Then in 2016, two independent groups did a search around the same time from the same set of 16 galaxy clusters from the HIFLUGCS sample with the largest $J$-factors, using five  and seven years of data respectively~\cite{Anderson16,Liang16}. The first analysis did not find any significant excess in the energy range between 10 to 400 GeV and upper limits on the velocity averaged annihilation cross-section  of around $3 \times 10^{-25} \rm{cm^3 s^{-1}}$ were obtained~\cite{Anderson16}.  The second analysis found a line at 43 GeV after stacking with a significance of $3.0\sigma$~\cite{Liang16}. However, since the backgrounds were not fully understood, they did not claim a detection and set a conservative limit on the the velocity averaged annihilation cross-section~\cite{Liang16}. A search for dark matter annihilation from galaxy groups (lower mass analogs of galaxy clusters) having large  $J$-factors was carried out using 413 weeks of Fermi-LAT data~\cite{Lisanti18}. No evidence for dark matter annihilations was found and   thermal relic cross sections for dark matter masses below  approximately 30 GeV  
to $b\bar{b}$ annihilation channel were excluded at 95\% c.l. Then,  another search for dark matter annihilation using 12 years of Fermi-LAT data from five nearby clusters (Centaurus, Coma, VIRGO, Fornax, and Perseus) reported null results and constraints on annihilation cross-section  into three different channels were reported, which are between $10^{-26}-10^{-23} ~\rm{cm^3 s^{-1}}$, depending on the WIMP mass~\cite{Thorpemorgan2021}. A search for line and box-shaped signals was done in ~\cite{Shen2021} using 7.1 years of Fermi-LAT data by  stacking the same set of 16 clusters analyzed in ~\cite{Anderson16,Liang16}. This work also found the 42 GeV line first reported  in ~\cite{Liang16}, although its significance was reduced. Consequently, 95\% c.l. upper limits on the thermally averaged  DM annihilation cross-section to photons between $10^{-27}-10^{-25} \rm{cm^3 s^{-1}}$ were set~\cite{Shen2021}. About a year ago,~\citet{DiMauro2023} (D23, hereafter) used 12 years of Fermi-LAT data to look for dark matter annihilation  and decay from 49 clusters selected from the HIFLUGCS sample. The stacking analysis  revealed a signal at 2.5-3.0$\sigma$ significance. However, the best-fit values of the mass and cross-section are in tension with the null results obtained using dark matter searches from  dwarf spheroidal galaxies. Therefore, it was concluded that the signal is most likely due to  cosmic ray collisions with the gas and photon fields within the cluster~\cite{DiMauro2023}.  Most recently, a constraint on the lifetime of  very heavy dark matter with masses between $10^3$ and $10^{11}$ GeV was set using 14 years of Fermi-LAT data, based on the analysis of seven clusters~\cite{Song2024}. 

Complementary to  the aforementioned target searches from individual clusters or a stacked analysis, a cross-correlation analysis between the 9-year Fermi-LAT diffuse gamma-ray map and four different galaxy cluster catalogs  was done to look for dark matter annihilations, which reported null results~\cite{Tan20}. In addition to galaxy clusters, a large number of works have also looked for dark matter annihilations and decay from Galactic center, dwarf spheriodal galaxies as well as Milky way satellites~\cite{Murgia}.

In this work, we look for  WIMP dark matter annihilation from  galaxy clusters detected by the  South Pole Telescope (SPT) 2500 square deg survey, which have been detected using the Sunyaev-Zeldovich (SZ) effect~\cite{SZ}, as a follow-up to our previous work which looked for point-source gamma-ray emission using 15 years of Fermi-LAT data~\cite{Manna2024}. We mainly follow the same methodology as D23. In our analysis, we have assumed a flat $\Lambda$CDM cosmology with  $\Omega_m = 0.3$ and $H_0 = 70  \text{\ } \text{km} \text{s}^{-1} \text{Mpc}^{-1}$.
This manuscript is structured as follows. In Section~\ref{sec:level3}, we define the cluster catalog utilized in our manuscript. Section~\ref{sec:level2} provides a comprehensive description of the halo and subhalo modelling. Subsequently, in Section~\ref{sec:level4}, we detail the analysis process employed in our study. Finally, in Section~\ref{sec:level5}, we present our results. The upper limits for all clusters can be found in Sect.~\ref{sec:level6}. We present our conclusions in Section~\ref{sec:conclusions}.

\section{\label{sec:level3}SPT-SZ cluster catalog\protect}
For our analysis,  we use the SZ selected cluster catalog detected by SPT, which provides a mass-limited sample.  The SPT  is a 10-meter telescope located at the South Pole that has imaged the sky at three different frequencies:  95 GHz, 150 GHz, and 220 GHz with an angular resolution of about 1 arcminute~\cite{Carlstrom}. SPT completed a 2500 square degree survey between 2007 and 2011 to detect galaxy clusters using the SZ effect. This survey detected 677  galaxy clusters with  SNR greater than 4.5, corresponding to a mass threshold of $3 \times 10^{14} M_{\odot}$ up to redshift of  1.8~\cite{Bleem15,Bocquet2019}\footnote{{\url{ https://pole.uchicago.edu/public/data/sptsz-clusters/2500d_cluster_sample_Bocquet19.fits }}}.   The SPT cluster redshifts were obtained using a dedicated optical and infrared follow-up campaign from pointed imaging and spectroscopic observations~\cite{Song,Ruel} in conjunction with  data from optical imaging surveys such as the Blanco Cosmology Survey~\cite{Desai12} and Dark Energy Survey~\cite{Saro}. We have previously done a search for gamma-rays induced due to astrophysical processes from 300 clusters from this sample, ranked according to $M_{500}/z^2$~\cite{Manna2024}.   Here, $M_{500}$ denotes the total mass enclosed within a sphere with a mean density of 500 times the critical density of the universe at the cluster's redshift, and $z$ represents the cluster's redshift~\cite{Bocquet2019}. In this work, we conduct a search for gamma-ray emission from dark matter annihilation for a sample of 350 galaxy clusters from the aforementioned catalog, after ranking them in decreasing order of their mass ratio, $M_{500}/z^2$. 

\section{\label{sec:level2}Dark Matter Analysis\protect}
The expected gamma-ray flux from the annihilation of WIMPs can be computed using the following equation~\cite{DiMauro2023,Charbonnier2012,Bonnivard2016,Hutten2019,Cirelli2011}\footnote{We follow the same notation as D23}: 
\begin{equation}
    \frac{d\Phi_\gamma}{dE}(E, \Delta\Omega, \text{l.o.s}) = \frac{d\phi_\gamma}{dE}(E) \times J(\Delta\Omega, \text{l.o.s}).
\end{equation}
Here, $\frac{d\Phi_\gamma}{dE} (E, \Delta\Omega, \text{l.o.s})$ is the WIMP dark matter induced differential gamma-ray flux per unit energy and solid angle as a function of energy ($E$), solid angle ($\Delta\Omega$), and line of sight (l.o.s.).  The term $\frac{d\phi_\gamma}{dE}(E)$ is the gamma-ray energy spectrum per annihilation as a function of energy and represents the particle physics component of the gamma-ray flux, which encapsulates the spectral characteristics of the WIMP annihilation such as the properties of the dark matter particles and the annihilation channels involved. 
The differential gamma-ray flux is then computed as follows~\cite{DiMauro2023}:
\begin{equation}
\frac{d\phi_\gamma}{dE}(E)= \frac{{\langle\sigma v\rangle}}{8\pi m_\chi^{2}} \times \frac{dN_\gamma(E)}{dE} .
\end{equation}
Here, $\frac{dN_\gamma(E)}{dE}$ denotes the WIMP photon spectrum, whose calculation is discussed in ~\cite{Cirelli2011}. The mass of the WIMP, denoted as $m_{\chi}$, determines the energy scale of the gamma rays produced during annihilation. The thermally-averaged annihilation cross-section, represented by $\langle\sigma v\rangle$, quantifies the likelihood of dark matter particles annihilating when they encounter each other. We can define the astrophysical $J$-factor $J(\Delta\Omega, \text{l.o.s})$ as the line-of-sight integral of the squared DM density profile over the solid angle at a given coordinate in the l.o.s. directions~\cite{Pace2019,DiMauro2023}. The $J$-Factor can be mathematically written as: 
\begin{equation}
 J(\psi,\theta, \Delta \Omega) = \int_{0}^{\Delta \Omega}\int_{\rm{l.o.s}} \mathrm{d}l \, \mathrm{d}\Omega \times \,\rho_{\rm tot}(r)^2 {\quad \rm }   
\end{equation}
Here, $\Delta\Omega$ is the solid angle over which the integration is performed, $\rho_{\rm tot}(r)$ is the total dark matter density profile as a function of the distance $r$ from the centre of the system, and the integration is carried out along the line-of-sight (l.o.s.). The solid angle $\Delta\Omega$ is related to the integration angle $\alpha_{\rm int}$ as:
\begin{equation}
   \Delta \Omega = 2 \pi (1 - \cos \alpha_{\text{int}}) \end{equation}
where $\alpha_{\rm int}$ is the angle between the line-of-sight and the direction pointing toward the centre of the cluster. We have chosen $\alpha_{int}$ to be 0.2$^{\circ}$ similar to size of the spatial bins used for the analysis.
In the context of modelling the dark matter distribution within galaxy clusters, a common approach is to consider the total dark matter density profile as the sum of two components~\cite{DiMauro2023}:
\begin{equation}
\rho_{\rm tot}(r) = \rho_{\rm main}(r) + \langle\rho_{\rm subs}\rangle(r)
\end{equation}
The first term on the right-hand side, $\rho_{\rm main}(r)$, represents the smooth, large-scale distribution of dark matter within the main halo of the galaxy clusters, whereas the second term 
($\langle\rho_{\rm subs}\rangle(r)$) accounts for the contribution of the population of subhalos  within the main halo of the galaxy cluster, according to the standard $\Lambda$CDM cosmological model~\cite{Prada2012,Klypin2011,Gao2004}. Subhalos are smaller-scale dark matter structures that are gravitationally bound within the larger main halo~\cite{Kuhlen2012,Zavala2019}. Since individual subhalos cannot be resolved, we use a statistical description of the subhalos to calculate the annihilation signal. 

In the next subsection, we discuss the modelling of  the smooth halo and subhalo dark matter component. This modelling process allows us to calculate the expected fluxes arising from DM annihilation within these clusters.

\subsection{Modelling the Main Halo}
\label{sec:NFW}
For our work, we modelled the main smooth dark matter halo using the Navarro-Frenk-White (NFW) density profile~\cite{Navarro1996,Navarro1997,Zhao1996}, which can be written as follows:
\begin{equation}
    \rho(r) = \frac{\rho_0}{\frac{r}{r_s}\left(1 + \frac{r}{r_s}\right)^2}
\label{eq:NFW}    
\end{equation}
Here, $\rho_0$ is the characteristic dark matter density for the NFW profile and $r_s$ is the NFW scale radius. The NFW profile is usually reparameterized in terms of the halo mass and concentrations as described below.

The first step in obtaining the halo concentration involves choosing a mass proxy. For this purpose, we use $M_{200}$, which is the cluster mass at a spherical overdensity, which is about 200 times the critical density of the Universe.
We have obtained  $M_{200}$ values from the SPT-SZ 2500 square-degree catalogue~\cite{Bocquet2019,Bleem2015}. In accordance with \cite{DiMauro2023,White2001}, we define the virial radius, $R_{200}$ which can be estimated from $M_{200}$ as follows:
\begin{equation}
 R_{200} = \left(\frac{3M_{200}}{4\pi\Delta_{200}\rho_{crit}}\right)^{1/3}   
\end{equation}
where $\Delta_{200}$ is the overdensity factor, which is 200 in this case. The critical density, $\rho_{crit}$, is calculated using the Hubble parameter $H(z)$, which is the rate at which the universe expands and is given by:
\begin{equation}
 \rho_{crit} = \frac{3H^2(z)}{8\pi G}
\end{equation}
where $H(z) \equiv H_0\sqrt{\Omega_M(1+z)^3+1-\Omega_M}$ for flat $\Lambda$CDM. We now introduce the  halo concentration ($c_{200}$) defined as $c_{200} \equiv \frac{R_{200}}{r_s}$.

Next, we need to determine  $c_{200}$ for every cluster. Although, an extensive weak lensing campaign has been undertaken for  SPT clusters for precise mass measurements and cosmological analyses~\cite{Bocquet2019} (and references therein),  no concentration measurements are available on a per cluster basis. Therefore, we use the  concentration-mass ($c-M$) relations from cosmological simulations for modelling the dark matter halo.  Similar to D23, we use the concentration-mass relations from ~\cite{SanchezConde2014}, which have been shown to match the observational results  across a wide range of halo masses from dwarf spheroidal galaxies to galaxy clusters. This mass-concentration relation allows us to obtain $c_{200}$ from the estimated $M_{200}$ for every SPT cluster.

The scale density $\rho_0$ in Eq.~\ref{eq:NFW} can be computed from $c_{200}$ as follows:
\begin{equation}
   \rho_0 = \frac{2 \Delta_{200} \rho_{\text{crit}} c_{200}}{3 f(c_{200})},
\end{equation}
where $f(c_{200})$ is given by:
\begin{equation}
    f(c_{200}) = \frac{2}{\left({c_{200}^2}\right)} \left[\ln(1+c_{200}) - \frac{c_{200}}{1+c_{200}}\right].
\end{equation}

\subsection{Modelling the subhalo}
\begin{figure}
\centering
\begin{adjustbox}{width=0.5\columnwidth}
\includegraphics[width=0.8\columnwidth]{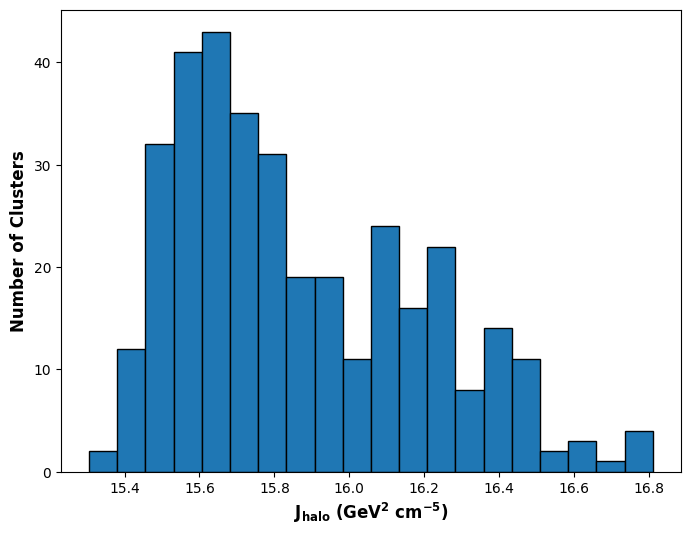}
\end{adjustbox}
\caption{Distribution of $J$-Factors for all the 350 SPT-SZ clusters used in our analysis.}
\label{fig:figure1}
\end{figure}

Structure formation follows a bottom-up hierarchical process in the standard $\Lambda$CDM cosmological model~\cite{Frenk2012}. This means that the smallest structures, known as subhalos, form first. Over time, these subhalos merge and accrete matter, leading to the formation of larger and larger structures, such as galaxy groups and clusters~\cite{Zavala2019}. 
Galaxy clusters are expected to host a large number of subhalos.
We shall now discuss the modelling of  the subhalo population along the same lines as D23. The subhalos account for the smaller-scale clumps and inhomogeneities within the halo. 
\\
For this purpose, we use the \texttt{CLUMPY v3} software~\cite{Charbonnier2012,Bonnivard2016,Hutten2018, Hutten2019}\footnote{{\url{ https://clumpy.gitlab.io/CLUMPY/v3.1.1/_downloads/975332ce6631f0956830ed27431f3b25/CLUMPY_v2018.06.CPC.tar.gz }}} where the Einasto profile \cite{Einasto1965} is used to describe the density distribution of subhalos.
The distribution of subhalos within a main halo is influenced by both their distance from the host center  and also their  mass. The total distribution of subhalos can be expressed as the product of three uncorrelated probability distribution functions (PDFs) and a normalization factor~\cite{DiMauro2023,Charbonnier2012,Bonnivard2016,Hutten2018} as follows:
\begin{equation}
    \frac{\mathit{d}^3 N}{\mathit{d}V\mathit{d}M\mathit{d}c} = N_{tot}\, \frac{\mathit{d}{\cal P}_V(R)}{\mathit{d}V}\, \times \frac{\mathit{d}{\cal P}_M(M)}{\mathit{d}M} \times \frac{\mathit{d}{\cal P}_c}{\mathit{d}c}(M,c) .
    \label{eq:clumpy}
\end{equation}
In the above equation, $N_{tot}$ refers to the expected number of subhalos within the virial radius of main halo, and $P_i$ with $i =  V, M, c$ is the probability distribution in each of the respective domains, normalized to 1. Here,  $V$  corresponds to the volume of the main halo, $M$  pertains to the distribution of subhalo masses, and $c$ represents the subhalo concentration.
Although, numerical cosmological simulations have significantly advanced our understanding of halo substructures~\cite{Zavala2019}, several questions still remain unresolved, such as the minimum mass for clump formation~\cite{Profumo05}, the effects of tidal stripping on subhalo survival~\cite{Aguirre2023}, and the exact shape of the subhalo DM density profiles~\cite{Errani2021}. These uncertainties impact the calculation of the DM-induced $\gamma$-ray flux. In this work, we follow the same prescription as D23 to account for each of the terms in Eq.~\ref{eq:clumpy}, which we describe below:
\begin{itemize}
    \item Spatial distribution PDF ($dP_v(R)/dV$): Since the main halo is spherically symmetric, the distribution of subhalos within it  depends only  on their distance from the central point of the host halo. It describes the probability of finding a subhalo at a given distance from the host halo centre and provides the spatial distribution of substructures in a host halo, which in our study are the galaxy clusters. We modelled it using the Einasto profile. 
    \item Mass distribution PDF ($dP_m(M)/dM$): Subhalo mass function can be defined as:
    \begin{equation}
        \frac{{\rm d}{\cal P}_M}{{\rm d}M} \propto M^{-\alpha}\,,
    \end{equation}
    We have used $\alpha$ = 1.9 in our study, similar to \cite{DiMauro2023,Springel2008}. 
    It represents a more conservative approach as it implies lesser number of subhalos. It also aligns with the results reported in \cite{Zavala2019}. 
    \item Concentration distribution PDF ($dP_c(M,c)/dc$): As reported in~\cite{Aguirre2023,Springel2008,Errani2020,Errani2021,Kazantzidis2004,Penarrubia2010}, subhalos experience tidal forces that generally result in significant mass loss, particularly in their outer regions. As a result, subhalos tend to be more concentrated compared to the  main halos with equivalent mass~\cite{Moline2017,Moline2023,Newton2022}. We have used ($c-M$) relation  defined in~\cite{Moline2017} for the subhalos, similar to D23. It accounts for the spatial dependence of the subhalos  within the main halos.
\end{itemize}
In order to account for the uncertainties in the above factors, D23 considered three benchmark models. In this work, we choose the parameters corresponding to the MED model in D23.  The smallest subhalo mass is assumed to be  \(10^{-6}\)$M_{\odot}$ and the largest subhalo mass is   0.01\% of the halo mass.  Finally, we specify the number of multilevel substructures to be two, as in D23. 
All the relevant properties used for modelling of dark matter halos and subhalos are summarized in Table~\ref{tab:TableI}.
The distribution of $J$-factors for SPT-SZ clusters using this model for the dark matter halo and substructure can be found in Figure~\ref{fig:figure1}.
The values  for individual clusters are listed in Table~\ref{tab:TableII}.

\section{\label{sec:level4}Fermi-LAT Analysis\protect}
\subsection{Data selection}
We conducted our search using 15.7 years of Fermi-LAT  data from August 5, 2008 to April 1, 2024, \texttt{(MET 239587201-733622405)}. For each cluster, a circular region of interest (ROI) with a 10$^\circ$ radius centered on the cluster was defined. The analysis was restricted to events within this ROI with energies ranging from 1 GeV to 300 GeV with 37 logarithmic energy bins per decade. We excluded lower energy events (below 1 GeV) due to the degraded  Point Spread Function (PSF) at those energies \cite{Ackermann2012,Ackermann2012b}. The binned maximum-likelihood method was implemented  using the \texttt{P8R3\_ULTRACLEANVETO\_V3} instrument response functions (IRFs) and \texttt{Pass 8 ULTRACLEANVETO ("FRONT+BACK")} class events \cite{Atwood2013}. 
\\
We also applied the filters, \texttt{DATA\_QUAL > 0} and \texttt{LAT\_CONFIG == 1}, which were utilized to eliminate low-quality data. The rock angle was constrained to \texttt{abs(rock angle) $< 52^\circ$} and \texttt{|b| $< 20^\circ$}  to mitigate particle contamination. To reduce atmospheric interference, a zenith angle cut of 90$^\circ$ was imposed on the events. We also removed all clusters within $20^\circ$ of the galactic plane. 
The spatial binning of the data was performed at a resolution of $0.2^{\circ}$ pixels, which helps to reduce noise and improve the signal-to-noise ratio. Our background model incorporated all the gamma-ray sources from the fourth Fermi-LAT catalogue \texttt{(4FGL-DR4)}~\cite{Ballet2023}, encompassing both point-like and extended sources. To account for the pervasive, low-intensity gamma-ray emission from the Milky Way galaxy itself (diffuse emission), we employed the Galactic diffuse emission model \texttt{(gll\_iem\_v07.fits)}. Additionally, an isotropic component \texttt{(iso\_P8R3\_ULTRACLEANVETO\_V3\_v1.txt)} is included to represent the isotropic gamma-ray background from various unresolved extragalactic sources. The analysis pipeline employed the  \texttt{Fermitools (v2.2.0)} software within the Fermi Science Support Center's Fermibottle Docker environment \cite{Wood2017}.  
\subsection{Analysis Method}
Our analysis followed the standard binned-likelihood method developed by the Fermi-LAT collaboration for likelihood calculations and model fitting~\cite{Atwood2009}. We incorporated sources from the 4FGL catalog up to $10{^\circ}$ beyond the defined region of interest into our model, freezing all their parameters to the catalog values. This was done to minimize bias from the potential presence of bright sources outside the targeted area and to account for the LAT's poor point spread function (PSF) at low energies around 0.1 GeV~\cite{Thorpemorgan2021,Manna2024}. In the first phase of the analysis, the spectral parameters for all the  free sources were determined.  During the second phase, we then fixed the spectral parameters of all sources, except for the normalizations to their best-fit values. Then, we included a template for potential dark matter annihilation  in the model,  which we modeled as a diffuse source with a spatial distribution proportional to the $J$-factor calculated using CLUMPY and specified in Table~\ref{tab:TableII}. 
The spectral component of the signal was given by the spectrum of annihilating dark matter for the specified channel utilizing the ~\texttt{DMFitFunction}~\cite{Jeltema2008} within {\tt Fermitools} similar to ~\cite{Thorpemorgan2021}.
We used the \texttt{gtapps} tool for our analysis and employed the Maximum Likelihood Estimation (MLE) technique to identify the model parameters that best match the source's spectrum and location. The \texttt{gtlike} tool performs a binned likelihood analysis on Fermi-LAT gamma-ray data. It achieves this by comparing a model of the gamma-ray sky with actual observations and calculating the likelihood that the model explains the data. To identify gamma-ray sources and assess their significance, we employ the Test Statistic (TS) calculated using the \texttt{gttsmap} tool. This statistic, defined as~\cite{Mattox1996} 
\begin{equation}
   TS = -2\ln(L_{max,0} / L_{max,1}), 
\end{equation}
compares the likelihoods of two scenarios: where $L_{max,0}$ is a model without the source (null hypothesis) and $L_{max,1}$ is a model including the source at a specific location (alternative hypothesis). 
 Wilks' theorem tells us that for high photon counts, the TS for the null hypothesis behaves similar to a chi-squared distribution~\cite{Wilks1938}.  The detection significance  is thereby nominally  obtained by taking the square root of TS. However, D23 has shown using simulations, that for the null hypothesis the TS statistics has larger tails, which are not expected in the $\chi^2$ distribution. Therefore, the actual detection significance would be expected to be smaller than the square root of TS. Nevertheless, we shall quote the significance as the square root of TS for our results.
We note that the  same TS statistic is also used in neutrino and soft gamma-ray astrophysics to evaluate the significance of detections~\cite{Pasumarti2022,Manna2024b}.
\begin{table}
    \centering
    \caption{\textbf{Parameters used for modelling the dark matter halo and subhalo.}}
     \label{tab:TableI}
    \begin{tabular}{|l|l|}
        \hline
        \textbf{Parameter} & \textbf{Value} \\
        \hline \hline
        Smallest subhalo mass & \(10^{-6}\)  $M_{\odot}$  \\
        Biggest subhalo mass & 0.01 \% of host mass \\
        Number of multilevel substructures & 2 \\
        Mass-concentration model for Halos & \cite{SanchezConde2014} \\
        Mass-concentration model for subhalos & \cite{Moline2017} \\
        \(dP/dV\) profile of subhalo distribution in host & EINASTO \\
        Slope of power law subhalo mass spectrum \(dP/dM\) & 1.9 \\
        Fraction of host halo mass bound in subhalos & 0.1 \\
        Branching Ratio Channel &  $b\bar{b}$,$\tau^+ \tau^-$ \\
        Spectrum Model & \cite{Cirelli2011} \\
        Dark Matter Particle & Majorana \\
        
        \hline 
    \end{tabular}
\end{table}

\section{\label{sec:level5}Results\protect}
\begin{figure}

\centering
\begin{adjustbox}{width=0.5\columnwidth}
\includegraphics[width=0.8\columnwidth]{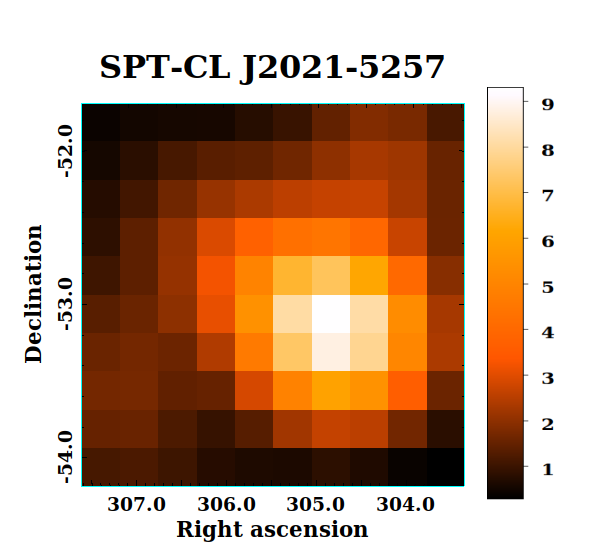}
\end{adjustbox}
\caption{Gaussian kernel smoothed ($\sigma = 1.5$) TS map of the SPT-CL J2021-5257 cluster (left) and TS map scale (right) generated using \texttt{gttsmap} in the energy band $1- 300$ GeV. We used 0.2-pixel resolution for the spatial binning.}
\label{fig:figure2}
\end{figure}

\begin{figure}
\centering
\begin{adjustbox}{width=0.5\columnwidth}
\includegraphics[width=0.8\columnwidth]{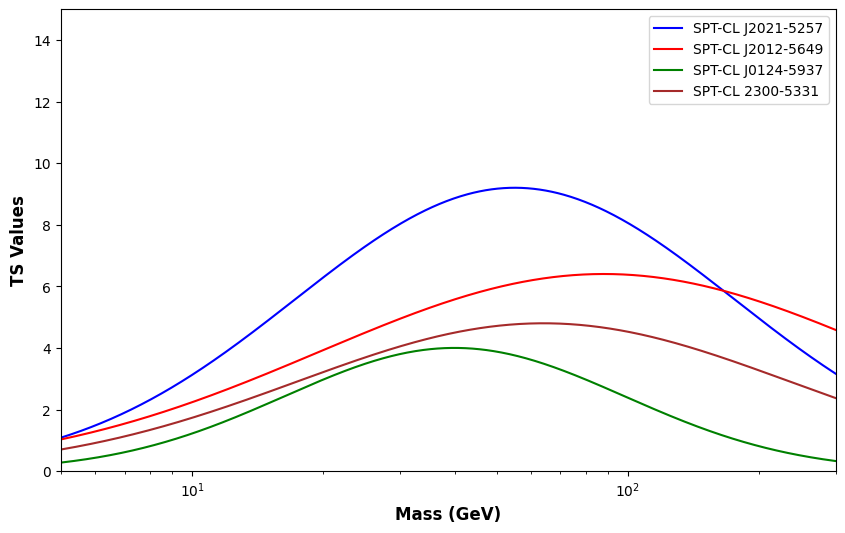}
\end{adjustbox}
\caption{TS as a function of the WIMP mass for clusters with TS $>=4$ for the  $b\bar{b}$ annihilation channel.}
\label{fig:figure3}
\end{figure}

We conducted the analysis for the clusters in our sample using the $b\bar{b}$ and $\tau^+ \tau^- $ annihilation channels in the energy range between 1-300 GeV.
The TS values for all the 350 galaxy clusters (for the  $b\bar{b}$ annihilation channel) can be found in Table~\ref{tab:TableII}.
Among all the clusters,  SPT-CL J2021-5257 shows the maximum TS value of 9.2, corresponding to 3.03$\sigma$ significance. The corresponding TS map for this cluster is  shown in Figure~\ref{fig:figure2}. We also performed the analysis for SPT-CL J2021-5257 by extending the  lower energy range to 100 MeV, and found that the  TS value decreases to 7 corresponding to 2.6$\sigma$ significance.
Other clusters, such as SPT-CL J2012-5649, SPT-CL J0124-5937, and SPT-CL J2300-5331 show TS values corresponding to $\sim 2-2.5\sigma$ significance, suggesting a weaker detection significance. The variation of TS values as a function of the dark matter mass for the aforementioned clusters with significance $>=2\sigma$ can be found in Figure~\ref{fig:figure3} for $b\bar b$ annihilation channel.
The best fit WIMP mass for SPT-CL J2021-5257 is found to be $(60.0 \pm 11.8)$ GeV, whereas the best-fit value of $\langle \sigma v \rangle = (6.0 \pm 0.6) \times 10^{-25} \text{cm}^3 \text{s}^{-1}$ for this cluster assuming a $b\bar{b}$ annihilation channel.  For SPT-CL J2012-5649, SPT-CLJ 0124-5937, and SPT-CL J2300-5331, the best-fit masses were found to be $(90.0 \pm 4.5)$ GeV, $(42.5 \pm 5.7) $ GeV, and $(65.0 \pm 3.8)$ GeV, respectively. The corresponding best-fit values of $\langle \sigma v \rangle$ for the $b\bar{b}$ channel were $(5.9 \pm 0.5) \times 10^{-24} \text{cm}^3 \text{s}^{-1}$, $(9.0 \pm 0.3) \times 10^{-25} \text{cm}^3 \text{s}^{-1}$ and $(3.5 \pm 0.2) \times 10^{-25} \text{cm}^3 \text{s}^{-1}$, respectively. The best-fit dark matter mass for  SPT-CL J2021-5257 for the $\tau^+ \tau^-$annihilation channel   is found to be $(15.3 \pm 4.1)$ GeV,  whereas the  best-fit annihilation cross-section is given by  $\langle \sigma v \rangle = (3.5 \pm 0.9) \times 10^{-25} \text{cm}^3 \text{s}^{-1}$.  Similarly, for SPT-CL J2012-5649, SPT-CLJ0124-5937, and SPT-CL J2300-5331 we found the best-fit mass as $(24.0 \pm 1.5)$, $(15.0 \pm 2.8)$, and $(22.0 \pm 2.1)$ GeV, respectively, whereas the   best-fit values of  $\langle \sigma v \rangle$ are equal to  $(5.0 \pm 0.4) \times 10^{-24} \text{cm}^3 \text{s}^{-1}$,  $(6.3 \pm 0.8) \times 10^{-25} \text{cm}^3 \text{s}^{-1}$ and $(2.8 \pm 0.4) \times 10^{-25} \text{cm}^3 \text{s}^{-1}$, respectively for this annihilation channel. The variation of TS value  with the WIMP mass   for  the $\tau^+ \tau^-$ annihilation channel for all the four clusters can be found in Fig.~\ref{fig:figure4}. We also show how TS varies with other subhalo models and uncertainties on $J$-factors in Appendix A, B, and C.

As pointed out in D23, these values are in conflict with the upper limits obtained from null searches for dark matter annihilation from Milky Way dwarf spheroidal galaxies, whose 95\% c.l. limits on the velocity-averaged annihilation cross-section are
around $10^{-26} \rm{cm^3/sec}$~\cite{Dimauro21}.  Therefore, although our significance does not cross the $5\sigma$ threshold, these enhanced TS values corresponding to (2-3)$\sigma$ significance cannot be by-products of dark matter annihilation. Furthermore, after considering the look elsewhere effect, given that we have analyzed 350 galaxy clusters~\cite{Gross}, the global $p$-value for the cluster with the highest TS value of around 9.2, is equal to 0.47, which is consistent with the expected  background. However, if we ignore the look elsewhere effect, and in case  the significance increases with additional exposure, then one explanation is that this signal could be due to some  astrophysical process resulting from cosmic interactions with gas and photons in the intracluster medium, as pointed out  in D23.

Therefore, we calculate upper limits for the dark matter annihilation cross-section for these clusters. For the aforementioned four clusters,  we show the 95\% c.l. upper limits on the annihilation cross-section as a function  of WIMP  mass for  the $b\bar{b}$ annihilation channel in Figure~\ref{fig:figure5} and for the  $\tau^+ \tau^-$ channel in Figure~\ref{fig:figure6}. Our upper limits  are mostly  in agreement with other works, which have obtained upper  limits on the  annihilation cross-section~\cite{DiMauro2023,Ackermann2010,Han2012,Thorpemorgan2021}. A compilation of upper limits for all the remaining  galaxy clusters analyzed in this work can be found in Table~\ref{tab:TableIII}, as discussed  in Sect.~\ref{sec:level6}. Among all the clusters with significance $>=2\sigma$, we found the most stringent limit for SPT-CL J2300-5331, viz. $\langle \sigma v \rangle  = 8.85 \times 10^{-26} \text{cm}^3 \text{s}^{-1}$ for $m_{\chi} = 10$ GeV considering the  $b\bar{b}$ annihilation channel, and $\langle \sigma v \rangle  = 10.0 \times 10^{-26} \text{cm}^3 \text{s}^{-1}$ for $m_{\chi} = 10$ GeV and considering the $\tau^+ \tau^-$ annihilation channel. We note that one of the clusters with $>=2\sigma$ significance is SPT-CL J2012-5649 (TS = 6.4), which is spatially coincident with the merging cluster Abell 3667.  This cluster was detected with the highest significance ($>6\sigma$) in our previous work on gamma-ray searches using a point source template~\cite{Manna2024}. 

All other clusters have significance less then $3\sigma$.
These include the merging clusters  in our sample such as  Bullet Cluster (SPT-CL J0658-5556) and El Gordo Cluster (SPT-CL J0102-4915)  with  TS values of 1.5 and 1.9, respectively, corresponding to no significant excess. Since the  last 100 clusters in our sample (sorted according to $M_{500}/z^2$) showed no significance we restricted our analysis to 350 clusters and did not go beyond that.

\hfill
\hfill
\begin{figure}
\centering
\begin{adjustbox}{width=0.5\columnwidth}
\includegraphics[width=0.8\columnwidth]{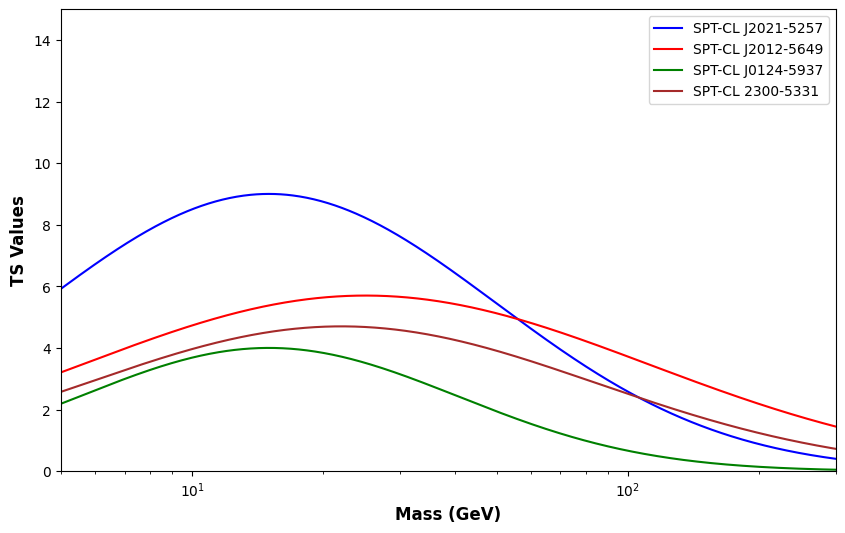}
\end{adjustbox}
\caption{Similar to Figure~\ref{fig:figure3} but here we show TS as a function of WIMP mass for the  $\tau^+ \tau^-$ annihilation channel.}
\label{fig:figure4}
\end{figure}
\hfill
\begin{figure}
\centering
\begin{adjustbox}{width=0.4\columnwidth}
\includegraphics[width=0.8\columnwidth]{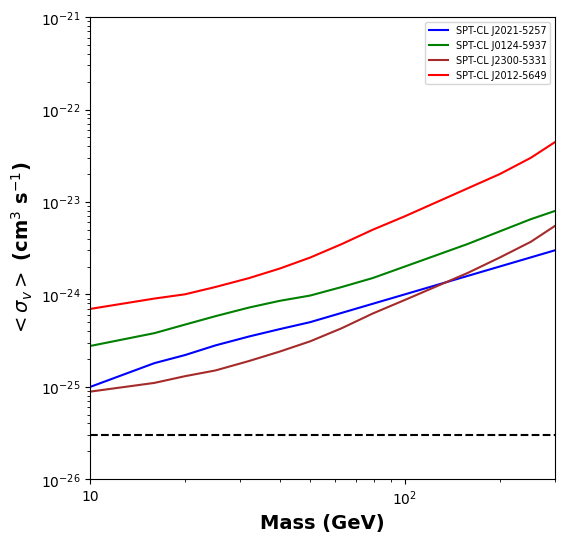}
\end{adjustbox}
\caption{95 $\%$ c.l. upper limits on the annihilation cross-section of WIMP dark matter for the \texttt{$b\bar{b}$} annihilation channel, considering the presence of substructures in the galaxy clusters with TS $>=4$. The black solid line indicates the canonical thermal cross-section of $3 \times 10^{-26} \, \text{cm}^3\text{s}^{-1}$.
}
\label{fig:figure5}
\end{figure}
\hfill
\begin{figure}
\centering
\begin{adjustbox}{width=0.4\columnwidth}
\includegraphics[width=0.8\columnwidth]{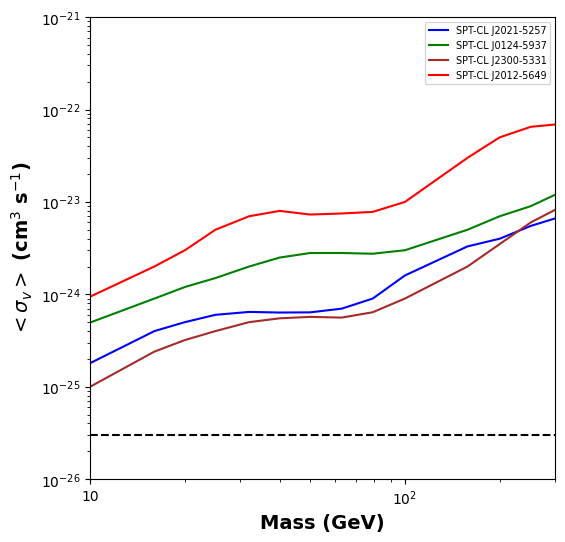}
\end{adjustbox}
\caption{Similar to Figure~\ref{fig:figure4} but here we show the 95\% c.l. upper  limits  for $\tau^+ \tau^-$ annihilation channel similar to~\cite{DiMauro2023}.}
\label{fig:figure6}
\end{figure}

\clearpage
\begin{center}
\begin{longtable*}{|>{\centering\arraybackslash}p{3.5cm}|>{\centering\arraybackslash}p{1cm}|>{\centering\arraybackslash}p{1cm}|>{\centering\arraybackslash}p{2cm}|>{\centering\arraybackslash}p{1.5cm}|>{\centering\arraybackslash}p{2cm}|>{\centering\arraybackslash}p{1.5cm}|>{\centering\arraybackslash}p{2cm}|>{\centering\arraybackslash}p{1.5cm}|}
\caption{\label{tab:TableII} The SPT-SZ galaxy clusters along with their coordinates, virial mass,  redshifts, normalization, where $\rho_s = \rho_0/4$, scale radius, $J$ values and TS Values for $b\bar{b}$ annihilation channel  . The clusters are ranked in decreasing order of their $M_{500}/z^{2}$ values.All clusters with TS $>=4$ are highlighted in bold. We have integrated out to the size of the spatial bin which is 0.2$^{\circ}$.}
\\
\hline
\endfirsthead

\multicolumn{9}{c}%
{{\bfseries \tablename\ \thetable{} -- continued}} \\
\hline
\endhead

\hline
\endfoot

\endlastfoot
\textbf{Cluster Name} & \textbf{R.A.} & \textbf{Dec.} & \textbf{$\mathbf{M_{200}}$} & \textbf{$\mathbf{z}$} & \textbf{$\mathbf{\rho_{s}}$} & \textbf{$\mathbf{r_{s}}$}  & \textbf{$\mathbf{J_{\text{halo}}}$} & \textbf{TS Values}
\\
&  ($^{\circ}$) & ($^{\circ}$) &  $  \times 10^{14} \, M_{\odot} / h$  &  &  $M_{\odot} \, \text{Kpc}^{-3}$ & (Kpc) & $\text{GeV}^{2} \, \text{cm}^{-5}$  &  \\ 
\hline \hline
\textbf{SPT-CL J2012-5649} & \textbf{303.11} &\textbf{-56.83} & \textbf{7.22} &\textbf{0.06}&\textbf{$\boldsymbol{2.97\times 10^{5}}$}&\textbf{416.3}  &\textbf{$\boldsymbol{6.45 \times 10^{16}}$} &\textbf{6.4} \\
SPT-CL J0431-6126 &67.84&-61.44&7.2 &0.06&$2.97\times 10^{5}$&416.0  &$6.00 \times 10^{16}$&1.5  \\
SPT-CL J2313-4243 &348.5&-42.73&6.71 &0.06&$2.98\times 10^{5}$&405.8  &$5.85 \times 10^{16}$&2.0  \\
SPT-CL J2009-4518 &302.45&-45.31&6.23 &0.06&$2.99\times 10^{5}$&395.2  &$5.25 \times 10^{16}$&1.9  \\
SPT-CL J2201-5956 &330.47&-59.94&14.03 &0.1&$2.92\times 10^{5}$&523.0  &$5.60 \times 10^{16}$&0.8  \\
SPT-CL J0328-5541 &52.17&-55.7&8.18 &0.08&$2.95\times 10^{5}$&435.0  &$4.55 \times 10^{16}$&3.5  \\
\end{longtable*}
\multicolumn{6}{l}{} \\[-1pt] 

\end{longtable*}
\end{center}
\section{\label{sec:level6}Upper Limits on annihilation cross-section for all clusters\protect}
We now calculate  the calculate upper limits on the annihilation cross-section for $b\bar{b}$ and $\tau^+ \tau^-$ annihilation channels for all the 350 clusters in our sample. These limits  are collated in Table~\ref{tab:TableIII}.   Amongst all the remaining clusters, we found the most stringent limit for SPT-CL J0455-4159, viz. $\langle \sigma v \rangle  = 6.44 \times 10^{-26} \text{cm}^3 \text{s}^{-1}$ for $m_{\chi} = 10$ GeV and $b\bar{b}$ annihilation channel.  The corresponding limit for $\tau^+ \tau^-$ annihilation channel for this cluster is given by   $\langle \sigma v \rangle  = 8.26 \times 10^{-26} \text{cm}^3 \text{s}^{-1}$ for $m_{\chi} = 10$ GeV.
\begin{center}
\begin{longtable*}{|>{\centering\arraybackslash}p {4cm}|>{\centering\arraybackslash}p {4cm}|>{\centering\arraybackslash}p {4cm}|}
\caption{\label{tab:TableIII} 95\% c.l.upper limits on $\langle \sigma v \rangle$ for all 350 clusters for $b\bar{b}$ and $\tau^+ \tau^-$ annihilation channels for WIMP mass of 10 GeV.}
\\
\hline
\endfirsthead

\multicolumn{3}{c}%
{{\bfseries \tablename\ \thetable{} -- continued}} \\
\hline
\endhead

\hline
\endfoot

\endlastfoot
\textbf{Cluster Name}  & \textbf{$b\bar{b}$ (\text{cm$^3$ s$^{-1}$})} & \textbf{$\tau^+ \tau^-$ (\text{cm$^3$ s$^{-1}$})}
\\
& &   \\
\hline \hline
SPT-CL J2012-5649 &  $6.95 \times 10^{-25}$  & $9.44 \times 10^{-25}$ \\
SPT-CL J0431-6126 &  $6.20 \times 10^{-25}$  & $8.10 \times 10^{-25}$ \\
SPT-CL J2313-4243 &  $5.97 \times 10^{-25}$  & $6.72 \times 10^{-25}$ \\
SPT-CL J2009-4518 &  $5.93 \times 10^{-25}$  & $8.14 \times 10^{-25}$ \\
SPT-CL J2201-5956 &  $3.14 \times 10^{-25}$  & $7.72 \times 10^{-25}$ \\
SPT-CL J0328-5541 &  $1.75 \times 10^{-25}$  & $6.54 \times 10^{-25}$ \\
\end{longtable*}
\end{center}

\subsection{Combined Upper limits from all clusters}
\label{sec:combined}
\begin{figure}
\centering
\begin{adjustbox}{width=0.5\columnwidth}
\includegraphics[width=0.8\columnwidth]{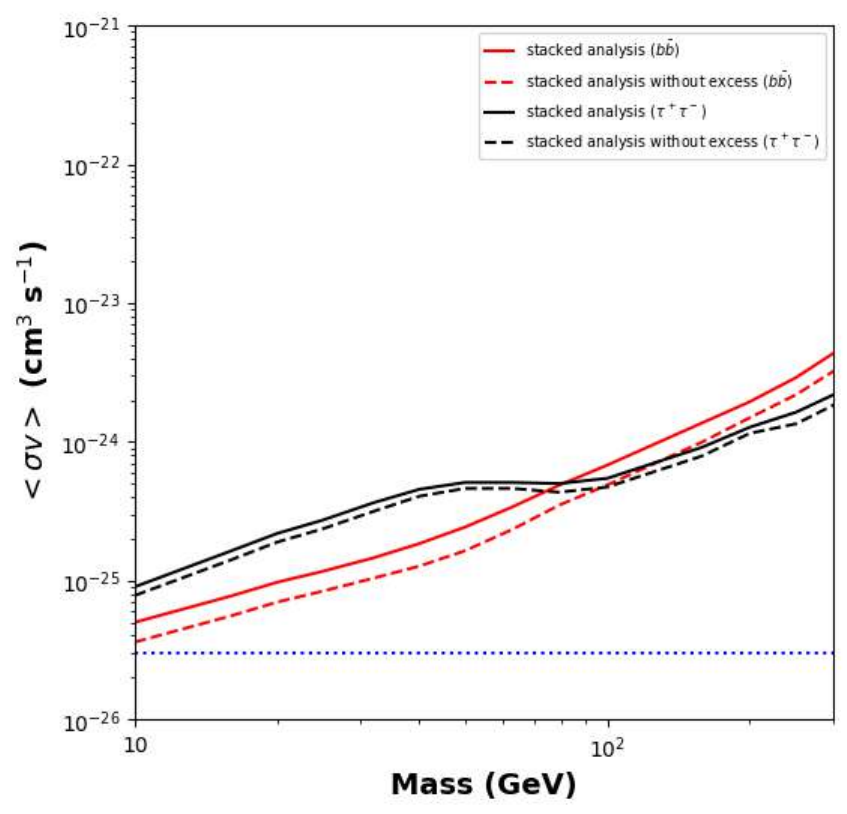}
\end{adjustbox}
\caption{95\% CL upper limits on the annihilation cross-section for the $b\bar{b}$ (red) and $\tau^+ \tau^-$ (black) annihilation channels using the combined data from all clusters. The solid lines represent the results for all 350 clusters, while dashed lines show the limits when clusters with TS $> 4$,  viz, SPT-CL J2012-5649, SPT-CL J2300-5331, SPT-CL J2021-5257 and SPT-CL J0124-5937 are excluded.  The dotted blue line shows the canonical thermal cross-section of $3 \times 10^{-26}$ cm$^3$ s$^{-1}$.}
\label{fig:figure10}
\end{figure}

We constructed the likelihood profiles for each individual cluster and then summed them to calculate a cumulative  limit on the  annihilation cross-section. This process is applied to all 350 clusters, both including and excluding those with TS $> 4$. We use the same methodology as described in~\cite{Huang11,McDaniel2024}. As shown in Fig.~\ref{fig:figure10}, we first include all 350 clusters and present the annihilation cross-section limits for the $b\bar{b}$ as well as  $\tau^+ \tau^-$ channel. Next, we exclude the clusters with TS $> 4$, viz, SPT-CL J2012-5649, SPT-CL J2300-5331, SPT-CL J2021-5257 and SPT-CL J0124-5937 and recalculate the upper limits which are superposed in Fig.~\ref{fig:figure10}. The inclusion of high-significance clusters introduces an excess signal, which somewhat biases the cross-section limits towards higher values. However, the combined limits are still roughly of the same order of magnitude. 

\section{\label{sec:conclusions}Conclusions\protect}
In this work, we searched for gamma-ray emission between 1 and 300 GeV from dark matter annihilation in galaxy clusters. We  analyzed 350 clusters selected from the SPT-SZ 2500 square degree survey, which provides a mass-limited sample. The analysis utilized 15.7 years of data from the Fermi-LAT telescope and employed the \texttt{DMFIT} template. Notably, we identified a $3\sigma$  signal from the galaxy cluster SPT-CL J2021-5257.  The best fit value of $\langle \sigma v\rangle$ for SPT-CL J2021-5257 is found to be $(6.0 \pm 0.6) \times 10^{-25}~\rm{cm^{3} s^{-1}}$ and the best fit WIMP mass  as $60.0 \pm 11.8$ GeV for the  $b\bar{b}$ annihilation channel.  For the  $\tau^+ \tau^-$ annihilation channel, we found the best -fit value as $\langle \sigma v\rangle = 3.5 \pm 0.9 \times 10^{-25} ~\rm{cm^{3} s^{-1}}$ and best-fit WIMP mass  as $15.3 \pm 4.1$ GeV. The TS map for this cluster can be found in Fig.~\ref{fig:figure2}. Although prima-facie, we found $3\sigma$ detection evidence from SPT-CL J2021-5257, the estimated annihilation cross-section is in conflict with upper limits from Milky way dwarf spheroidal galaxies.
Therefore, we  conclude that the enhanced signal seen for SPT-CL J2021-5257 cannot be due to dark matter annihilation. Furthermore, if we consider the look elsewhere effect the global $p$-value for this cluster is consistent with the expected background.
We also obtained a marginal significance of  around 2$\sigma$  for three other clusters: SPT-CL J2012-5649, SPT-CL J0124-5937, and SPT-CL J2300-5331. All the remaining clusters showed null results and the TS values were consistent with background.

Therefore, we then calculate upper limits for all the analyzed clusters.
Among the clusters with $>2\sigma$ significance we found the most stringent limit for SPT-CL J2300-5331, viz. $\langle \sigma v \rangle  = 8.85 \times 10^{-26} \text{cm}^3 \text{s}^{-1}$ for $m_{\chi} = 10$ GeV and $b\bar{b}$ annihilation channel and $\langle \sigma v \rangle  = 10.0 \times 10^{-26} \text{cm}^3 \text{s}^{-1}$ for $m_{\chi} = 10$ GeV and $\tau^+ \tau^-$ annihilation channel at  95\% c.l.  Among all the remaining clusters,  the most stringent  upper limits for annihilation to $b\bar{b}$ is obtained for  SPT-CL J0455-4159, as $\langle \sigma v \rangle  = 6.44 \times 10^{-26}~\text{cm}^3 \text{s}^{-1}$ for $m_{\chi} = 10$ GeV for  $b\bar{b}$ annihilation channel, and $\langle \sigma v \rangle  = 8.26 \times 10^{-26}~ \text{cm}^3 \text{s}^{-1}$ for $m_{\chi} = 10$ GeV for  $\tau^+ \tau^-$ annihilation channel at 95\% c.l. {We also show the combined upper limits for all the clusters in Fig.~\ref{fig:figure10}. In  the Appendices, we also check the robustness of our results by redoing the analysis with  different assumptions for the  dark matter substructure  and log-normal uncertainties on the $J$-factor.


Moving forward, we plan to expand this analysis to encompass all galaxy clusters detected in ongoing X-ray and SZ surveys such as eROSITA~\cite{erosita} and SPTPol~\cite{Bleem24}. 

\begin{acknowledgments}
We are grateful to the Fermi-LAT and CLUMPY teams for their invaluable contributions in making the data and analysis codes publicly available. Their expertise and willingness to answer our queries were instrumental in this research. This work was motivated after a stimulating talk by Mattia Di Mauro at IIT Hyderabad. We also thank the anonymous referee for useful constructive feedback on our manuscript.
We also acknowledge the National Supercomputing Mission (NSM) for providing essential computing resources on the `PARAM SEVA' supercomputer at IIT Hyderabad. This initiative is implemented by C-DAC and supported by the Ministry of Electronics and Information Technology (MeitY) and the Department of Science and Technology (DST) of the Government of India. Finally, we thank the Smithsonian Astrophysical Observatory for developing SAOImageDS9, which played a valuable role in our analysis. 
\end{acknowledgments}
\bibliography{references}

\appendix
\section{Boost Factor using different subhalo models}
\label{sec:Boost Factor}
In addition to calculating the boost factors using the subhalo model discussed  in the main manuscript, we followed the conservative models for the subhalo  outlined in~\cite{Bartles2015} and the more aggressive model defined in~\cite{Gao2012}, and calculated the boost factor as a function of minimum subhalo mass for all the four clusters with significance $>2\sigma$. The conservative model in ~\cite{Bartles2015} uses two different {\it ansatzes} for the subhalo mass function obtained from ~\cite{SanchezConde2014}.
The first subhalo mass function considered is given by   $\frac{dN}{dm} = \frac{0.03}{M} \left(\frac{m}{M}\right)^{-1.9}$, where $\frac{dN}{dm}$ describes the number density of subhalos per unit mass \textit{m} within a host halo of mass \textit{M}. For SPT-CL J2021-5257,  we obtained a $J$-factor of $2.30 \times 10^{16}$ and a TS value of 8.5 using this model. The second subhalo mass function used in ~\cite{Bartles2015} is given by  $\frac{dN}{dm} = \frac{0.012}{M} \left(\frac{m}{M}\right)^{-2.0}$, for which the  $J$-factor increased to $2.15 \times 10^{17}$ and the TS to 10.7 for this cluster.  We then considered the  aggressive model for the subhalo profile~\cite{Gao2012}, which adopted the results from high-resolution simulations of galaxy cluster haloes (the Phoenix Project). Using this aggressive model,  we found a $J$-factor of $3.13 \times 10^{18}$ and a TS of 11.5 for SPT-CL J2021-5257. These $J$-factor values correspond to a minimum subhalo mass of $10^{-6} M_{\odot}$. We  then redid the same exercise for three other clusters with significance $>2\sigma$ and report the $J$-Factors and TS values in Table~\ref{tab:Table IV}. The boost factors as a function of the minimum halo mass can be found in Figures~\ref{fig:figure7}, ~\ref{fig:figure8} and~\ref{fig:figure9} respectively.
Despite using these different boost factors, we do not find any cluster with TS$>25$. 

\begin{table}[h!]
    \centering
    \caption{\label{tab:Table IV}$J$-Factors and TS values for four clusters (with TS$>4$) using three different methods for subhalo mass distributions and $b\bar{b}$ annihilation channel. The first two columns represent the conservative models described in~\cite{Bartles2015}, and the other one is a more aggressive model defined in~\cite{Gao2012}.}
    \begin{tabular}{|>{\centering\arraybackslash}p {4cm}|>{\centering\arraybackslash}p {2cm}|c|>{\centering\arraybackslash}p {2cm}|c|>{\centering\arraybackslash}p {2cm}|c|}
        \hline
        \multirow{2}{*}{Cluster} & \multicolumn{2}{c|}{~\cite{Bartles2015} with Power Law Index -1.9} & \multicolumn{2}{c|}{~\cite{Bartles2015} with Power Law Index -2.0} & \multicolumn{2}{c|}{~\cite{Gao2012}} \\ \cline{2-7}
                                 & $J$-Factor & TS & $J$-Factor & TS & $J$-Factor & TS \\ 
                                 & \text{[GeV$^2$ cm$^{-5}$]} & & \text{[GeV$^2$ cm$^{-5}$]} & & \text{[GeV$^2$ cm$^{-5}$]} & \\ \hline
        SPT-CL J2021-5257 & $2.30 \times 10^{16}$  & 8.50 & $2.15 \times 10^{17}$ & 10.70 & $3.13 \times 10^{18}$  & 11.50 \\ \hline
        SPT-CL J2012-5649 & $5.90 \times 10^{16}$ & 6.00 & $2.10 \times 10^{17}$ & 7.20 & $4.70 \times 10^{18}$ & 7.70 \\ \hline
        SPT-CL J0124-5937 & $1.42 \times 10^{16}$ & 3.70 & $4.77 \times 10^{16}$ & 4.55 & $2.35 \times 10^{18}$ & 5.00 \\ \hline
        SPT-CL J2300-5331 & $1.18 \times 10^{16}$ & 4.35 & $7.56 \times 10^{16}$ & 5.65 & $1.77 \times 10^{18}$ & 6.20 \\ \hline
    \end{tabular}
\end{table}

\begin{figure}
\centering
\begin{adjustbox}{width=0.5\columnwidth}
\includegraphics[width=0.8\columnwidth]{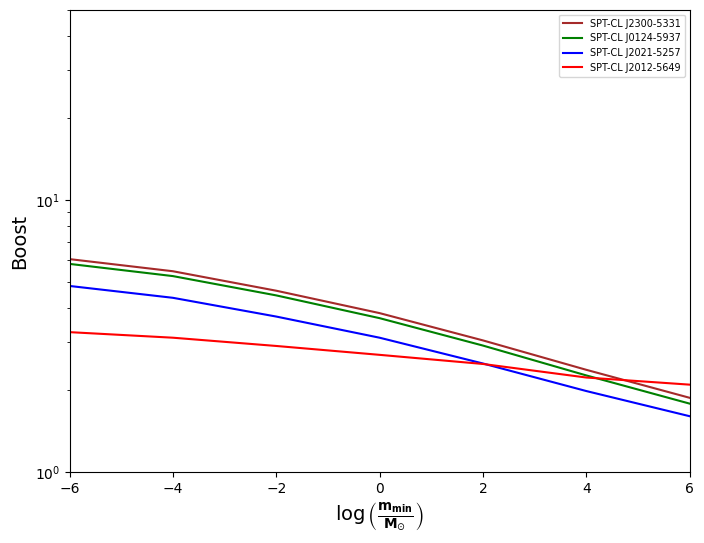}
\end{adjustbox}
\caption{The boost as a function of the minimum subhalo mass for all four clusters with significance>2$\sigma$ using the subhalo
mass function from~\cite{SanchezConde2014} with Power Law Index -1.9. The methodology followed is similar to~\cite{Bartles2015}.}
\label{fig:figure7}
\end{figure}
\hfill
\begin{figure}
\centering
\begin{adjustbox}{width=0.5\columnwidth}
\includegraphics[width=0.8\columnwidth]{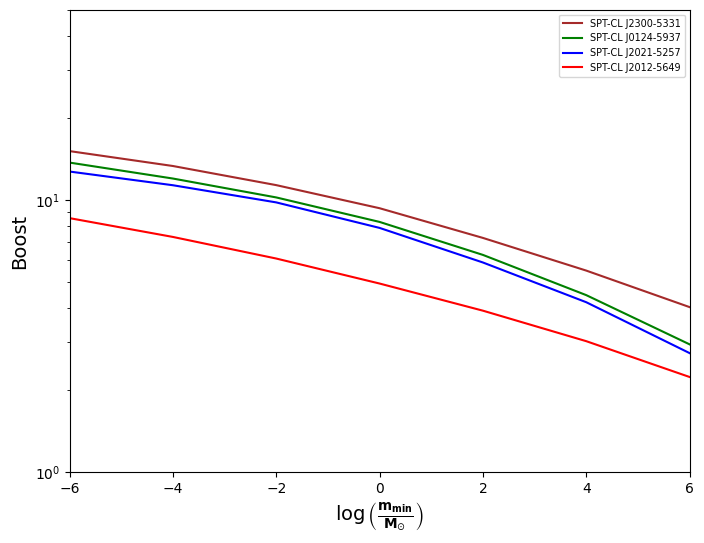}
\end{adjustbox}
\caption{Same as Fig.~\ref{fig:figure7} but here we use power law index as -2.0.}
\label{fig:figure8}
\end{figure}
\hfill
\begin{figure}
\centering
\begin{adjustbox}{width=0.5\columnwidth}
\includegraphics[width=0.8\columnwidth]{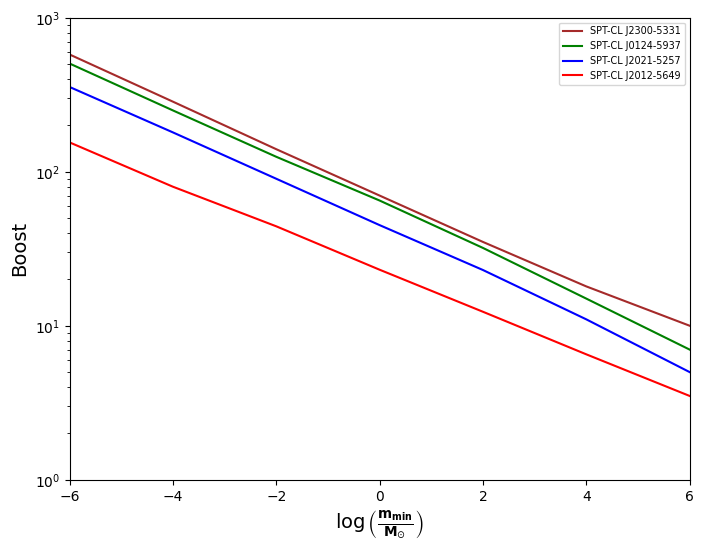}
\end{adjustbox}
\caption{Same as Fig.~\ref{fig:figure7} but here we followed the subhalo model described in~\cite{Gao2012}.}
\label{fig:figure9}
\end{figure}
\section{Best-Fit WIMP parameters using different subhalo models}
\label{sec:upper_limits}
We  now recalculate the  best-fit values of mass and annihilation cross-section using the conservative subhalo models defined in~\cite{Bartles2015} and the aggressive model in ~\cite{Gao2012}, as discussed in Appendix~\ref{sec:Boost Factor}. 
The best-fit values for the WIMP mass and annihilation cross-section for all the four clusters with TS$ > 4$ can be found in Table~\ref{tab:Table V} for both $b\bar{b}$  and $\tau^+ \tau^-$   annihilation channels.
All the best-fit values for the annihilation cross-section are $\mathcal{O} (10^{-25}-10^{-24})~ \rm{cm^3s^{-1}}.$
Therefore, the new   best-fit values are still in conflict with the upper limits from dwarf spheroidal galaxies~\cite{Dimauro21}. 
Among the clusters with TS $> 4$, the lowest value for the best-fit annihilation cross-section was obtained for SPT-CL J2300-5331 using the aggressive model defined in~\cite{Gao2012}. For the $b\bar{b}$ annihilation channel, this best-fit value is equal to  $\langle \sigma v \rangle = (9.3 \pm 0.4) \times 10^{-26} \, \text{cm}^3 \text{s}^{-1}$, with a dark matter particle mass of $m_{\chi} = (52.6 \pm 3.7)$ GeV. For the $\tau^+ \tau^-$ annihilation channel, the corresponding best-fit cross-section is $\langle \sigma v \rangle = (8.8 \pm 0.2) \times 10^{-26} \, \text{cm}^3 \text{s}^{-1}$, with a WIMP mass of $m_{\chi} = (18.8 \pm 2.5)$ GeV. ~\citet{Dimauro21} obtained  a 95\% upper limit for the annihilation cross-section of dark matter particles on Dwarf spheroidal galaxies, of $\langle \sigma v \rangle = 1.5 \times 10^{-26} \, \text{cm}^3 \text{s}^{-1}$ for the $b\bar{b}$ annihilation channel for WIMP mass of $m_{\chi} = 50$ GeV. 
For the $\tau^+ \tau^-$ annihilation channel, the corresponding 95\% upper limit is $\langle \sigma v \rangle = 9 \times 10^{-27} \, \text{cm}^3 \text{s}^{-1}$  for WIMP mass  of $m_{\chi} = 20$ GeV.  Therefore, the lowest values of the annihilation cross-section which we obtain using the most aggressive halo models are still in tension with the upper limits from dwarf spheroidal galaxies  at the  best-fit  WIMP mass which we found.

\begin{table}[h!]
    \centering
    \caption{\label{tab:Table V} The best-fit mass and annihilation cross section for four clusters with TS $>4$ for both $b\bar{b}$ and $\tau^+ \tau^-$ annihilation channels using three different models. The first two columns represent the conservative models described in~\cite{Bartles2015}, and the other one is a more aggressive model defined in~\cite{Gao2012}.}
    \begin{tabular}{|>{\centering\arraybackslash}p {3cm}|>{\centering\arraybackslash}p {2cm}|c|>{\centering\arraybackslash}p {2cm}|c|>{\centering\arraybackslash}p {2cm}|c|}
        \hline
        \multirow{2}{*}{Cluster} & \multicolumn{6}{c|}{$b\bar{b}$ Channel} \\ \cline{2-7}
                                 & \multicolumn{2}{c|}{~\cite{Bartles2015} with Power Law Index -1.9} & \multicolumn{2}{c|}{~\cite{Bartles2015} with Power Law Index -2.0} & \multicolumn{2}{c|}{~\cite{Gao2012}} \\ \cline{2-7}
                                 & $m_x$ & $\langle \sigma v \rangle$ & $m_x$ & $\langle \sigma v \rangle$ & $m_x$ & $\langle \sigma v \rangle$  \\ 
                                 & (GeV) & (\text{cm$^3$ s$^{-1}$}) & (GeV) & (\text{cm$^3$ s$^{-1}$}) & (GeV) & (\text{cm$^3$ s$^{-1}$}) \\ \hline
        SPT-CL J2021-5257 & $78.3 \pm 7.5$ & $(2.4 \pm 0.5) \times 10^{-24}$ & $55.8 \pm 4.5$ & $(5.1 \pm 0.4) \times 10^{-25}$ & $48.3 \pm 4.7$ & $(2.5 \pm 0.9) \times 10^{-25}$ \\ \hline
        SPT-CL J2012-5649 & $105.8 \pm 11.1$ & $(8.0 \pm 0.7) \times 10^{-24}$ & $80.3 \pm 3.7$ & $(4.7 \pm 0.5) \times 10^{-24}$ & $72.8 \pm 5.1$ & $(9.8 \pm 0.5) \times 10^{-25}$ \\ \hline
        SPT-CL J0124-5937 & $62.3 \pm 3.4$ & $(2.7 \pm 1.3) \times 10^{-24}$ & $35.7 \pm 6.1$ & $(5.5 \pm 0.7) \times 10^{-25}$ & $29.0 \pm 3.5$ & $(3.0 \pm 0.7) \times 10^{-25}$ \\ \hline
        SPT-CL J2300-5331 & $73.9 \pm 4.7$ & $(6.1 \pm 0.8) \times 10^{-25}$ & $58.0 \pm 4.2$ & $(2.1 \pm 0.3) \times 10^{-25}$ & $52.6 \pm 3.7$ & $(9.3 \pm 0.4) \times 10^{-26}$ \\ \hline
        
        \multirow{2}{*}{Cluster} & \multicolumn{6}{c|}{$\tau^+ \tau^-$ Channel} \\ \cline{2-7}
                                 & \multicolumn{2}{c|}{~\cite{Bartles2015} with Power Law Index -1.9} & \multicolumn{2}{c|}{~\cite{Bartles2015} with Power Law Index -2.0} & \multicolumn{2}{c|}{~\cite{Gao2012}} \\ \cline{2-7}
                                 & $m_x$ & $\langle \sigma v \rangle$ & $m_x$ & $\langle \sigma v \rangle$ & $m_x$ & $\langle \sigma v \rangle$  \\ 
                                 & (GeV) & (\text{cm$^3$ s$^{-1}$}) & (GeV) & (\text{cm$^3$ s$^{-1}$}) & (GeV) & (\text{cm$^3$ s$^{-1}$}) \\ \hline
        SPT-CL J2021-5257 & $27.4 \pm 2.3$ & $(7.8 \pm 1.2) \times 10^{-25}$ & $13.4 \pm 2.5$ & $(2.5 \pm 0.4) \times 10^{-25}$ & $10.7 \pm 1.3$ & $(1.8 \pm 0.3) \times 10^{-25}$ \\ \hline
        SPT-CL J2012-5649 & $33.1 \pm 1.8$ & $(7.2 \pm 0.6) \times 10^{-24}$ & $21.7 \pm 0.8$ & $(3.7 \pm 0.5) \times 10^{-24}$ & $19.5 \pm 1.8$ & $(7.5 \pm 0.5) \times 10^{-25}$ \\ \hline
        SPT-CL J0124-5937 & $24.0 \pm 3.1$ & $(1.8 \pm 1.0) \times 10^{-24}$ & $12.9 \pm 2.1$ & $(4.5 \pm 0.7) \times 10^{-25}$ & $11.0 \pm 1.5$ & $(2.7 \pm 0.6) \times 10^{-25}$ \\ \hline
        SPT-CL J2300-5331 & $25.7 \pm 3.0$ & $(5.3 \pm 0.7) \times 10^{-25}$ & $20.3 \pm 2.7$ & $(1.1 \pm 0.3) \times 10^{-25}$ & $18.8 \pm 2.5$ & $(8.8 \pm 0.2) \times 10^{-26}$ \\ \hline
    \end{tabular}
\end{table}

\section{Statistical uncertainty in the $J$-factor}
\label{sec:statistical_uncertainity}
In our earlier analysis,  we had not included the uncertainty in the central halo $J$-Factor. To account for this, we include a log-Gaussian uncertainty similar to Eq. 7 in~\cite{McDaniel2024} and Eq. 16 in D23. We include the statistical uncertainty on the $J$-factor by multiplying the Fermi-LAT binned Poisson likelihood function with a $J$-factor likelihood function that takes the form of a Gaussian in $\log_{10} (J)$ with width $\sigma_i$ as described below: 
\begin{equation}
L_i (J_i \mid J_{\text{obs},i}, \sigma_i) = \frac{1}{\log(10) J_{\text{obs,i}} \sqrt{2\pi} \sigma_i } \times \exp\left[ - \left(\frac{\log_{10}(J_i) - \log_{10}(J_{\text{obs,i}})}{\sqrt{2} \sigma_i}\right)^2 \right]
\end{equation}
The best-fit observed $J$-factor for the \textit{i}$^{th}$ cluster is denoted by $J_{\text{obs},i}$, and the error in $\log_{10}(J_{\text{obs},i})$  is denoted by $\sigma_i$.
 The value of the $J$-factor for which the likelihood is computed is represented by $J_i$. We 
considered three different values of $\sigma_i$, viz. 0.2, 0.4, and 0.6  dex similar to D23 and \cite{McDaniel2024}. For galaxy clusters, the uncertainty due to substructure and mass-concentration relation  has been estimated to be 0.2 in D23, whereas for dwarf galaxies, it is about 0.6 dex~\cite{McDaniel2024}. D23 also did the analysis for $\sigma_i=0.4$  dex. Therefore, we considered all the three values of $\sigma_i$. Nevertheless, D23 has shown that the best-fit values for the WIMP  mass and annihilation cross-section are not affected by $\sigma_i$.

We then proceed to calculate TS values for all the four clusters with TS values $>4$, after  accounting for log-Gaussian uncertainty in the $J$-Factors  using all the three values of $\sigma_i$ and report them in Table~\ref{tab:TableVI}. The results for TS are not much different compared to the earlier values and  we still do not find any cluster with TS $> 25$. There is also not much difference in the new TS values compared to the previous values which did not assume any uncertainty on $\sigma_i$. 

\begin{table}[h!]
\centering
\caption{\label{tab:TableVI} $J$-Factors including statistical uncertainty and TS values for clusters with significance > $2\sigma$ for the  $b\bar{b}$ annihilation channel.  We show the values for three different values of $\sigma_i$. For comparison we have also shown the J-Factors and TS values without including the uncertainty.}
\begin{tabular}{|l|c|c|>{\centering\arraybackslash}p{1.5cm}|c|}
\hline
\textbf{Cluster Name} & \textbf{log$_{10}$(J)  without $\sigma_i$} & \textbf{TS Values without $\sigma_i$} & \textbf{$\sigma_i$} & \textbf{TS Values} \\ 
                      & \text{[log$_{10}$GeV$^2$ cm$^{-5}$]} & & [dex] & \\ \hline
\multirow{3}{*}{SPT-CL J2021-5257} & \multirow{3}{*}{16.40} & \multirow{3}{*}{9.00} & 0.20 & 9.35 \\ \cline{4-5} 
                                   &                        &                       & 0.40 & 9.70 \\ \cline{4-5} 
                                   &                        &                       &0.60 & 10.20 \\ \hline
\multirow{3}{*}{SPT-CL J2012-5649} & \multirow{3}{*}{16.79} & \multirow{3}{*}{6.20} & 0.20 & 6.85 \\ \cline{4-5} 
                                   &                        &                       & 0.40 & 7.10 \\ \cline{4-5} 
                                   &                        &                       & 0.60 & 7.20 \\ \hline
\multirow{3}{*}{SPT-CL J0124-5937} & \multirow{3}{*}{16.26} & \multirow{3}{*}{4.00} & 0.20 & 4.15 \\ \cline{4-5} 
                                   &                        &                       & 0.40 & 4.50 \\ \cline{4-5} 
                                   &                        &                       & 0.60 & 4.73 \\ \hline
\multirow{3}{*}{SPT-CL J2300-5331} & \multirow{3}{*}{16.14} & \multirow{3}{*}{4.60} & 0.20 & 4.77 \\ \cline{4-5} 
                                   &                        &                       & 0.40 & 5.00 \\ \cline{4-5} 
                                   &                        &                       & 0.60 & 5.30 \\ \hline
\end{tabular}
\end{table}


\end{document}